# Scientometric Rules as a Guide to Transform Science Systems in the Middle East & North Africa

Jamal El-Ouahi[1,2]

[1] j.el.ouahi@cwts.leidenuniv.nl - https://orcid.org/0000-0002-3458-7503
Centre for Science and Technology Studies (CWTS), Leiden University, Leiden, Netherlands

[2] Clarivate Analytics, Dubai Internet City, Dubai, United Arab Emirates

**Abstract**
This study explores how scientometric data and indicators are used to transform science systems in a selection of countries in the Middle East and North Africa. I propose that scientometric-based rules inform such transformation. First, the research shows how research managers adopt scientometrics as 'global standards'. I also show how several scientometric data and indicators are adopted following a 'glocalization' process. Finally, I demonstrate how research managers use this data to inform decision-making and policymaking processes. This study contributes to a broader understanding of the usage of scientometric indicators in the context of assessing research institutions and researchers based on their publishing activities. Related to these assessments, I also discuss how such data transforms and adapts local science systems to meet so-called 'global standards'.

**Keywords**
Scientometric Rules, Research Evaluation, Research Policy, Science Systems, Middle East and North Africa

## 1. Introduction

Since the 1990s, research institutions have become a rapidly expanding research domain (Berman, 2011; Brisset-Sillon, 1997; Brunsson et al., 2012; Dearlove, 1997; Krücken & Meier, 2006; Musselin, 1996, 2005, 2013; Renaut, 1995; Siow, 1995; Van Vught, 1995). This trend reflects the fact that the performance of academic institutions is increasingly being scrutinised in light of its effects on economic growth and social equality as well as to address the demand for accountability from various stakeholders. In this context, tools inspired by New Public Management such as indicators, policies and rankings contribute to the vision that there is only one way to manage and evaluate quality in scientific research. As reported by Franssen and Wouters (2019), bibliometric methods have been extensively developed and employed in the context of science policy as a tool for research evaluation (Moed et al., 1985; Narin, 1976). Margolis (1967) presented an early use of citations to evaluate science. In his paper, Margolis was already discussing the 'new scale of values based on citations' as 'new standards'. These standards could be referred to as 'scientometric standards'. Despite their broad scope, bibliometric methodologies are currently mainly understood in the context of evaluative research management. Research management and evaluation have a considerable impact on knowledge production through the promotion of researchers, economic incentives, funding, and reputation. Some of the characteristics that are likely to influence research management and research evaluation are variations in research practices and publication strategies, as well as agreement on research objectives (Whitley & Gläser, 2007).

Scientific research has traditionally been evaluated primarily based on scientific papers, which constitute science's most visible and measurable output (Geuna & Martin, 2003; Hicks, 2012). Academics and research institutions are evaluated and ranked based on a variety of publishing performance criteria (Hirsch, 2005; Narin & Hamilton, 1996), which involves the allocation of research funds as well as the assignment of academic roles (Geuna & Martin, 2003; Hicks, 2012). A massive literature has focused on the accuracy of modern management and performance metrics, such as productivity, citation indexes, and peer review (Anninos, 2014; Basu, 2006; Werner, 2015). For example, H-indices, citation counts, and Journal Impact Factors (JIF) are bibliometric indicators widely used when evaluating research (Thelwall et al., 2015). There is a heterogenous literature about the formalized uses of metrics in research assessment (de Rijcke et al., 2016).

According to Weingart (2005), the introduction of bibliometric techniques is a response to the pressures on science systems to legitimate themselves. Various scientometric methods and indicators have been developed in the last 30 years and used by research managers and policymakers for various purposes such as institutional reporting but also to support the development of research directions and research policies (Ellegaard & Wallin, 2015; Moed et al., 1992). The use of scientometric data by research managers and policymakers might be seen as part of a shift to managerialism and increasing levels of accountability in research institutions (Langfeldt et al., 2020). For instance, several authors studied the management of research activity through the use of output performance measures such as the number of citations and the number of articles published in peer-reviewed journals (Agyemang & Broadbent, 2015; Osterloh, 2010). The use of scientometric data in research management and evaluation has been studied extensively (Jiménez-Contreras et al., 2003; Lahtinen et al., 2005; Morris et al., 2011; Sivertsen, 2018; Thomas et al., 2020). This use includes decisions in research evaluation such as faculty promotion or hiring.

So far, the use of scientometric data by policymakers and research managers in emerging nations, such as countries in the Middle East and North Africa (MENA), has received little attention. This study explores how research managers in MENA adapt 'global standards' to alter organizational research governance practices. The following questions are addressed in this study based on interviews with research managers:

- How do research managers in MENA adopt global scientometric standards in local contexts?
- How is this adoption implemented at the organizational level?
- In which local processes are scientometric data and indicators used and what specific functions do they serve?

This paper aims to provide a better understanding of research institutions as organizations, particularly with regard to how local management deals with global standards (Peterson, 2007). This study explores the complex interplay between local and global factors in shaping the practices of research managers. More specifically, it develops the notion of 'scientometric rules' that are set in local contexts to define and operationalize research quality scientometrically. By examining the local uses of scientometrics by research managers, the paper sheds light on the development of these scientometric rules as glocalized versions of 'global standards'.

This study is organized as follows. First, the relevant theories and concepts are introduced. Then, the methods and data used in the study are described. After that, the adoption and implementation of scientometrics as 'global standards' at the local level are discussed. Finally, the role of research managers in using scientometrics to make decisions and set science policies is examined, and potential future research opportunities are discussed.

## 2. Theories

This study draws upon the following theories and concepts that will be developed hereafter: judgment devices, objectivity, global standards and glocalization.

*2.1 Judgment devices and objectivity*

It is essential to comprehend the concepts of judgment devices and objectivity to understand how research managers use scientometric data in different contexts. These concepts play a crucial role in determining how scientometric data is used by research managers. An analogy can be made between the evaluation of research objects and the valuing of unique goods termed *singularities* by Karpik (2010). Singularities are goods that are unique and not difficult to compare to others, such as a novel, a work of art, a researcher or a scientific journal. The need for external assistance arises from the difficulty in evaluating singularities. Customers, or in our case research managers, rely upon external support in the form of judgment devices that help validate their judgments. According to Karpik (2010), judgment devices can be divided into five types: appellations, cicerones, confluences, networks, and rankings. Appellations and rankings are useful to understand the role of scientometrics when they are used in the context of research valuations.

Appellations are brands or titles names that assign a meaning or a certain value to a specific product or a group of products. *Nature* or *Science* are examples of such brands. Similarly, it could also be the indexation of a journal in a specific scientometric database such as the Web of Science. As per Karpik (2010), appellations or brands build on shared conceptions regarding the quality of a specific product. In instances where a quality agreement is not reached, another option is to make use of rankings. Rankings order singularities in a sorted list based on one or multiple criteria. Karpik makes the distinction between two types of rankings: those built on expert rankings, such as public rankings of universities by domain specialists, and those that make use of buyers' choices of a particular object, such as top 1% or top 10% cited publications in their fields.

In the book 'Trust in Numbers: The Pursuit of Objectivity in Science and Public Life', Porter (1996) explores the question of how to explain the prestige and influence of quantitative approaches in modern society. The analytical nature of scientometrics as a judgment device makes it the preferred tool to rationalize organizational management (Porter, 1996). He examines the development of the concept of objectivity in science and public life and how the quest for objectivity has influenced the evolution of social, political, and scientific institutions as well as how it has come to be a significant aspect of modern scientific culture. By conceptualizing quantification as a 'technology of distance', Porter specifically emphasizes the applicability of number-based devices for communication beyond the boundaries of locality. For instance, university rankings reduce complexity and can act as a link between the academic sector and other sectors. University management may perceive rankings as a useful tool to compare the performance of their institution with others, set some strategic goals or monitor the overall academic activity.

*2.2 Global Standards and glocalization*

The concepts of global standards and glocalization also provide a foundation for analyzing how scientometric data is used by research managers. In their book titled 'A World of Standards',

Brunsson and Jacobsson (2002) explore the concept of standards and how they function in modern society. The authors argue that essential aspects of contemporary society, such as norms, shape our lives in a variety of ways. They discuss the different standards, such as organizational, professional, and technical standards. Their book looks at the standard-setting process and how different entities, such as governments and institutions, influence how standards are developed and adopted. Their analysis shows that standard-setting is a multi-actor, dynamic process with a range of interests, objectives, and resources. Power dynamics play a role in this process, allowing dominant parties to shape the creation and adoption of standards to meet their objectives. As actors' requirements and interests change, standards can be altered, updated, or replaced (Brunsson et al., 2012). Brunsson and Jacobsson also cover the processes that create uniformity between organizations, and more specifically the diffusion of standards, their innovation or imitation.

The uniformity of academic institutions has been studied by Paradeise (2016) who explored whether higher education and research systems were in the process of becoming similar. She looks at how so-called 'global standards' affect academic institutions, in 'search of academic quality'. According to Paradeise (2016), there is a growing demand for academic institutions to follow global standards to ensure quality and boost their competitiveness. Her study focuses on the conflicts that arise in academic institutions between regional norms and global standards, as well as the difficulties that institutions encounter in balancing these two forces. She examines the process of developing and implementing standards in academic institutions, as well as how different actors, including governments, accrediting bodies, and international organizations, influence this process. Paradeise (2016) found that the globalization of academic activities and the world standards, such as performance rankings, tend to align the local level of quality to these standards. Paradeise and Thoenig (2013) explored the impact of 'global standards' on what academic quality means locally. Several arguments highlight a convergence among nations and universities with regard to higher education and research. Global standards of excellence, for instance, the importance of so-called A-ranked journal publications and citation indexes, have gained importance in recent years (Durand & Dameron, 2011). Paradeise and Thoenig (2013) also argue that ranking bodies, steering and evaluation bodies are predominantly in charge of controlling the definition of academic quality and assessing it. This paper examines the use of scientometric data by research managers and how they consider such data as 'global standards'.

The availability of scientometrics to order research objects hierarchically based on their performance creates some demand from research managers. Such information is considered useful by research managers in improving their organizations. This trust in numbers allocates a certain authority to scientometrics with which various research stakeholders engage locally, within their science systems or their institutions as a neutral, unbiased criterion. The localization of the scientometric used on a global scale to a level that matches the characteristics of the locality has been coined by Robertson (2012) as the *glocalization* process which describes how the local and the global interact to shape culture and society. It implies that, despite greater global connection and homogenization, globalization has also given local cultures and identities new opportunities to make themselves known. According to Robertson, local and global forces interact, transforming one another to produce a hybrid form of cultural expression and social organization that is both universal and local. Instead of supporting the idea that cultural homogenization results from globalization, this idea questions it and contends that it might result in a dynamic, multi-layered manifestation of cultural and social variety. In the context of this study, this means the adaptation of scientometric data into local markets. Scientometrics, which everyone can use, may be customized to conform to local preferences. This also supports the argument that local orders are still important, and that global standardization does not necessarily eliminate diversity (Paradeise & Thoenig, 2013). Glocalized scientometric indicators would be of much greater interest to the different research stakeholders because their localization makes it more specific to their context and their needs. A variety of judgment devices are used in different contexts, which motivates the exploration of how scientometric data and indicators are adopted, adapted and used by research managers in different national science systems.

## 3. Methods

Qualitative research is particularly useful to study topics where there is little literature such as the usage of bibliometrics in MENA. In this study, I adopt a qualitative approach based on interviews with 12 research managers working in 9 countries. An interpretive coding process was used to provide the results presented in the findings section. As set by Elo and Kyngäs (2008), several phases (*preparation*, *organizing,* and *reporting*) were followed.

In terms of *preparation*, the transcripts of the interviews constitute the empirical materials of this study. At the time of the interviews, all the research managers were affiliated with research-intensive universities located in MENA. Interviewing research managers from various countries is motivated by the wish to study various types of institutional settings, contexts, and backgrounds. 20 research managers were invited to participate in the study through an e-mail invitation explaining the purpose of the study, its main topics, and the expected duration of the interview. These research managers were selected from my professional network. The 20 research managers consist of current and ex-Clarivate customers, people I spoke to at scientometric conferences, and those I met in scientometric courses or online via Linkedin.

Among these 20, 12 responded positively. The interviews were conducted in English, French, and/or Arabic. They were conducted online with the exception of two face-to-face interviews. Table 1 lists the respondents by country along with the region and language(s) of the interview. In one instance (Respondent #12), the respondent was joined by a colleague from the same team. Before each interview, the interviewees were asked for permission to record the interview and to use anonymously the transcripts in a publication. They all gave their consent.

**Table 1. Country, region, and interview language of the respondents**

| Respondent | Country | Region | Language |
|---|---|---|---|
| 1 | Saudi Arabia | MENA | English/Arabic |
| 2 and 3 | Turkey (2 institutions) | MENA | English |
| 4 and 5 | Egypt (2 institutions) | MENA | English/Arabic |
| 6 | Iran | MENA | English |
| 7 | Tunisia | MENA | French |
| 8 and 9 | Morocco (2 institutions) | MENA | French/Arabic |
| 10 | Jordan | MENA | English/Arabic |
| 11 | Pakistan | MENA | English |
| 12 | Iraq | MENA | English/Arabic |

Interviews were semi-structured based on questions that allowed the interviewees to describe the contexts in which scientometric data is used in their institution. Some respondents also described the research assessment process at the national level. Open questions were asked about several dimensions of the usage of scientometric data:

a) data sources and data processing
b) responsibilities of the research manager
c) use cases of scientometric data
d) challenges and opinions

Additional questions were also asked about the organization of the institution's management and the decision processes. All interviews were conducted by the author and lasted one hour on average. Then, they were also transcribed verbatim by the author. Quotes originally in French or Arabic language and displayed in the findings section were translated into English by the author.

In the next phase of the content analysis, the collected data was *organized* as follows. In this iterative procedure, the transcripts were carefully read and openly coded. Then, they were reviewed, and the emergent categories were grouped into more precise categories. Finally, subcategories were grouped and refined hierarchically under subcategories that fall under the "main category" such as "scientometric uses". Several metrics or tools are not scientometric indicators *per se*, for example, journals rankings, a list of indexed journals, or a list of highly cited papers or highly cited researchers. But they are often derived from them. For that reason, such derived tools are also considered in this study. A broad definition of "scientometric use" was adopted in the early stages of the coding process. Then, such a definition was narrowed down and limited to situations in which they are used to make decisions or to set policies in various contexts. The reporting phase of this analysis does not include instances of categories such as "simple papers count". There are also examples of implicit or common knowledge of certain scientometric data which might be considered too broad to draw conclusions about their explicit uses in different contexts.

Finally, in the *reporting* phase, I provide quotes that illustrate the most prevalent subcategories that were formed during the analysis process to illustrate the findings. Based on the sample size and the various aspirations of the different research institutions, there might be some risk of overstating the presence of scientometric indicators in the context of decision or policymaking. The significance of the qualitative approach is in revealing how scientometrics are used in MENA to transform local science systems. This allows us to make conclusions and develop new research questions about the role and use of scientometric indicators in research management and evaluation which are briefly discussed in the last section of this paper.

## 4. The use of scientometrics as an implicit adoption of 'glocal' standards

*4.1 Scientometrics adopted as 'glocal' standards*

In this section, I show how using scientometric data is a first step in adopting global standards to alter local research systems. Scientometrics have been introduced in the 1960s (see Garfield (2009) for a brief history of the field). Since then, there have been significant advancements in technology, which are reflected in the ways that scientific information is communicated, retrieved, and evaluated (Garfield, 2012). At a conference in Istanbul (Turkey) in 2012,

Garfield mentioned how the Science Citation Index transformed from large and printed book volumes into the Web of Science to illustrate such evolution. The interest in the Web of Science also generated competition from Google Scholar, Scopus and other bibliographic and scientometric platforms which made access to citation data wider than before. The research managers under study argue that, before the introduction of scientometrics, there was very little information of this type they could retrieve about science.

> We are highly in favour of scientometrics. I remember in Chemistry, we didn't really know if somebody cited us. When I saw the Chemical Abstracts and the Chemical citations index, I was very impressed. So, without this information, it's almost impossible to evaluate someone abroad, even someone from the same University. It was embarrassing in Turkey. You should contact someone to know how many publications they had. But after the citation indexing was made available, this information was not private anymore. 30 years ago, in Turkey, you didn't have such visibility. Now everything is public. You cannot just say anything. Before, nobody cared about rankings. But 30 years ago, you wouldn't dare. I wouldn't dare myself rank countries or universities. (Respondent #3, Turkey)

In that sense, research managers claim that publications and citations data give them more information about science as well as a specific lens through which to view scientific knowledge production that was not available before the introduction of scientometrics. According to this research manager, this type of information also provides the capability to assess objects of evaluation with quantitative indicators such as the number of publications of a researcher and the number of citations these publications received. By explaining that because now everything is public one cannot just say anything, the research manager points to the origin of these indicators outside of the researchers themselves. Because citation databases are produced by companies and scientometric indicators by experts, these numbers are understood to represent a particular set of global standards that research managers connect to (Paradeise, 2016; Paradeise & Thoenig, 2013). This quote does not say anything about previous forms of research assessment other than that they did not have access on this scale to this type of information. Scientometrics offers a new lens through which research managers can assess science, research institutions, and researchers from a more distant, even global, perspective that allows one to rank countries and universities, something the manager did not dare to do 30 years earlier.

Scientometrics are also used to set certain sets of standards and rules to assess research objects. This is explicitly mentioned in the following quote, where the research manager relies on scientometric data to evaluate how valuable or reputed emerging journals are.

> It [scientometrics] is useful for Science. Because many different publishers, many different journals, emerge every day from different parts of the World. So, it's good to have some rules and some platforms to check scientometric data and to know how valuable these journals are. […] And we track the predatory journals on a list, every month. […] We use both Web of Science and Scopus to identify the journals. Many journals charge people some fees for publications, and they are not reputed journals. We check how many articles they have published, and how many times they have been cited. We would also check which databases index them. Scientometrics helps people how to publish their information better. (Respondent #6, Iran)

Hence, for this manager, what is to be considered a valuable and reputable journal, is defined by scientometric rules, based on the number of papers and citations of this journal as well as its indexation in certain bibliographic databases. Some research managers justify the usefulness and the adoption of scientometrics as a tool to distinguish the bad from the good, based on such scientometric rules. These rules are adopted and used as objective lenses to analyze scientific research at various levels.

An example of scientometric indicators used by some research managers are indicators at the researcher level but also indicators calculated at the journal level such as the journal impact factor quartile, which is calculated as the quotient of a journal's rank in its category and the total number of journals in a particular subject category. The quartiles rank the journals from highest to lowest based on their journal impact factor. By definition, there are four quartiles: Q1, Q2, Q3 and Q4.

> I follow some self-developed rules. There is no book to go by. I look at 2 main indicators. The first type of indicators is called the leading indicators which evaluates the quality of the journals a researcher published in. These include the quartile of a journal based on the impact factor. The papers published in high-impact factor journals will most likely get more citations. The citations and the Impact Factor are related. They go in the same direction. The second type of indicators is called the lagging indicators. I use the typical indicators at the researcher level: H-index, total citations, number of citations per paper, and the category normalized citation impact from InCites. I cannot wait for 4 or 5 years, for the publication to get cited to evaluate the researcher. That's why I go to the leading indicators. From an administration of research perspective, this is important for us. (Respondent #1, Saudi Arabia)

Here, the research manager follows some self-developed scientometric rules to evaluate the scientific publications of a researcher based on different types of indicators. There is a clear distinction made between the indicators defined at the journal level where one researcher has published and the indicators calculated at the researcher level. Due to citation latency, the journal indicators provide some useful information to the research manager that researcher-level indicators lack. These various scientometric indicators are used as a set of standards or rules to assess the quality of journals as a proxy of a researcher's standing.

Scientometric rules such as these show affinity with the role of quantification in the rise of modern society as argued by Porter (1996). By replacing human judgment in scientific communities and public life, quantification is understood to be more trustworthy and more objective (Karpik, 2010). Eventually, the individual indicators related to the journals that research managers use in such context are also understood or considered as global standards (Paradeise, 2016; Paradeise & Thoenig, 2013). They serve a specific function to define, represent and discuss research quality among the different research stakeholders.

*4.2 Implementation of 'glocalized' scientometric standards through teaching and negotiation processes*

The global rules represented in and through scientometrics are not self-evident but have to be actively communicated to researchers to be followed. Therefore, in this second section, I describe how the research managers convey the scientometric rules within their institutions through teaching and negotiation processes. Crucially, what I see here is that global standards are translated into local versions. This can be understood as a process of glocalization (Robertson, 2012) which localizes the scientometric used on a global scale to a level that fits the local needs. By adapting the scientometric data and indicators to local conditions, research managers aim at finding a balance between global standardization and localization.

Scientometrics is taught at different institutional levels. Many respondents explain that they conduct workshops and training sessions dedicated to researchers on how to analyze scientific research from the scientometric lens. For example, in the context of publication strategy, a research manager recommends to her management to frequently train the researchers of her institution on scientific publishing. The topics covered in these workshops range from scientific writing and publication to journal indexation across several databases and their scientometric-based rankings.

> I recommend that every 2 months, we should conduct workshops on how to publish, where to publish, how to choose journals, what are the best journals for each subject… We cover the following questions: What is an indexed journal? What is the indexing process? What is the difference between Web of Science and Scopus? We explain the impact factor and other metrics. We conduct small workshops and make them relevant to each subject category. First of all, the indexing of the journal is an important factor. The journal must be indexed. We explain what the Impact Factor, the Eigenfactor and the quartile of the journal are. (Respondent #4, Egypt)

Research managers also teach scientometrics to researchers as a set of rules. As a result, researchers also adopt scientometrics the way it is taught locally to them. In this case, teaching scientific publishing from a scientometric perspective is the main topic of these training sessions. This includes the coverage of several matters, such as the selection of the *best* journals in each subject for publication as well as the explanation of scientometric indicators.

Such teaching is also done for other various implicit purposes where research managers consider local variations in research practices and objectives. For example, in the next quote, the research manager explains that, as a consequence of teaching scientometrics across her institution, the number of papers of her institution indexed in the Web of Science grew, which also improved the global ranking of her institution in Essential Science Indicators, which take into account highly cited publications. She also mentions that her main role was to make sure that researchers published their work in *reputable* journals:

> When we started, there was none of these scientometric talks at the University. And there were some people, who had some information, and who wanted to do something but they didn't know where to start. So, we helped them to start […] Many researchers were very happy with this because they would see that the rank of the university is improving in the Web of Science and Essential Science Indicators. […] Making sure people are publishing their papers in reputable journals was our main daily routine job (Respondent #6, Iran)

The local translation of scientometrics is illustrated in the following quote where the research manager compares a particular researcher's publications in Q4 journals to those of the World, his country of affiliation, and his institution, for evaluation purposes.

> As a researcher, you would like to publish in high-impact factor journals. The higher the impact factor, the better the quality of the journal. It indicates the type of research we do. So, this researcher has 3% of his papers published in Q4 journals. The share of papers published in Q4 is 13% in the World, 10% in the country and 19% in the University. This researcher is not doing bad research. (Respondent #1, Saudi Arabia)

The first percentage mentioned in this quote (3% in Q4), and retrieved by the research manager, is the proportion of papers of the assessed researcher published in Q4 journals over a specific period. Then, the proportions of publications of the same Web of Science category (i.e. the field of the assessed researcher) published in Q4 journals are retrieved by the research manager at different levels: world, researcher's country of affiliation & researcher's affiliation. Such proportions are then considered as benchmarks. In the field of the evaluated researcher, the proportion of publications in Q4 journals over the same period is 13% at the world level, 10% at the country level, and 19% for the researcher's university. The research manager concludes that the concerned researcher is not doing *bad* research when comparing the proportion of his publications in Q4 journals with the proportion of publications in Q4 journals at the world level, the researcher's country of affiliation level and the researcher's university level.

Here, the research manager uses the Journal Impact Factor and the Quartile as an indicator of academic quality. Publishing frequently in Q4 or low-impact journals in their categories is interpreted as doing *bad* research. Such information shows that, in terms of publication strategy, the quartile is used as an obvious indicator of a journal's quality but also to evaluate the standing of a researcher in terms of research quality. Such use of glocalized scientometrics is meant to bridge the gap between the global research communities and local research stakeholders.

As mentioned earlier, the adoption of scientometrics at the institution and researcher level for publication strategy implies their communication to researchers by research managers. However, some research managers struggle with the introduction of scientometric indicators and the complex adoption of citation metrics by researchers in their institutions as stated in this quote:

> Another complicated challenge for us is, how to convince researchers that the indicators that are provided by the major databases are the right ones. For example, half of my researchers do not believe in the impact factor. They tell me, "That's not important. I'm not going to work under such pressure". That's a challenge, it's hard to prove to them that it's the best solution … although we are ranked in the national or global rankings, for me, it's hard to make it clear that it is important. For them a publication is good, it is not good or bad because the impact factor is this much or that much[1]. (Respondent #9, Morocco)

This quote highlights a significant 'challenge' faced by this research manager when trying to 'convince' researchers about the use of scientometrics as new standards for evaluation. On the one hand, this research manager clearly mentions the challenges he faces to prove to the researchers that the scientometric indicators are the *right* indicators to use and the *best* solution to choose the *right* or *good* journal for publication. This quote might suggest that some research managers are not entirely convinced by the use of scientometrics. On the other hand, this quote also shows that researchers have their own opinion about the use of scientometrics when valuing research quality. This quote implies that the adoption of scientometrics as glocal standards requires a negotiation process between research managers and researchers.

This section suggests that the teaching of scientometrics occurs in a glocal manner by incorporating them as global standards while also tailoring them to local needs and contexts. This teaching consists of presenting various indicators, such as journal indicators for publication or evaluation purposes, but also interpreting the same indicators to define 'quality' and 'reputation' (Paradeise, 2016; Paradeise & Thoenig, 2013). The quotes show that research managers have different practices and use glocal scientometrics to manage and evaluate research in their institutions. This section also suggests that scientometrics are communicated to researchers by research managers via a negotiation process as the *right* indicators or objective standards to evaluate the quality of journals for publication. This negotiation process involves recognizing and addressing the concerns and perspectives of researchers, as well as considering local variations in research practices. By adopting a glocal approach, research managers tailor metrics to local contexts and engage in a collaborative negotiation process with researchers to communicate the value and the uses of scientometrics.

## 5. Decision and policy making by research managers

*5.1 Hiring and promotion practices*

Scientometric indicators serve as glocalized rules and standards. The glocalization process previously discussed occurs also partly through the development of new decision-making processes as the scientometric indicators can be viewed as aids to facilitate and make decisions. In this sub-section, I demonstrate how research managers use scientometrics to hire and promote faculty members. Scientometric indicators serve the whole function to discuss research quality but also to make judgments about researchers (Karpik, 2010).

In the next quote, the hiring process of a faculty member is briefly explained. The research manager looks at the number of articles of the candidate. There is also a promotion process which involves the analysis of the number of papers published in journals indexed in three citation indices of the Web of Science Core Collection. This promotion process is a points-based system and assigns points to publication based on their types but also their indexation in the Web of Science:

> When the University wants to hire any new Faculty Member, the first thing we look at is the number of articles. There is also an official threshold, to become an associate professor. The faculty member needs to have at least some publications in the so-called "indexed journals", meaning indexed in Science Citation Index Expanded, Social Science Citation Index or in Arts Humanities Citation index. This is the background of all the promotion processes. If you publish a conference paper, it has a smaller value. If you publish a book in Turkish, it has 0 points. An academic book in Turkish does not mean anything. But if you publish in an indexed journal, this becomes more relevant. (Respondent #2, Turkey)

Scientometric indicators serve as decision-making devices and support a variety of research-related decisions. Research managers also use scientometrics to develop new goals and new policies through which the glocalization process of scientometric data discussed earlier also occurs. This implies the setting of new rules and organization goals which constitute a very direct form of implementation of scientometrics in science systems. The use of scientometric data is embedded in organizational processes which tend to copy the systems used in the United States as clearly stated in the quote below:

> Before, you had to apply to become a Special Professor and this title becomes valid all over Turkey. They had jury members. In the very beginning, starting from the 1930s until 1982 or something like that, you had to write another thesis in addition to your PhD thesis. This was ridiculous. And the jury would say ok this could work, and you have to enter another examination and answer questions about your field. And then you would need to give a lecture, so they can see how you lecture. Then, the Higher Education Council was established, and it was decided a Research Professor title should be awarded just like in the American system. They look at the number of articles, citations, and the number of theses you directed. It is still going that way. They appoint a jury before they see you.

> They look at your articles, and citations, and then they say let's take him to an oral exam. Recently, they eliminated the oral exam 2 years ago. Now they look at articles and citations only. (Respondent #3, Turkey)

This quote shows a clear transformation of the professorship promotion process. This transformation consists of several phases over a period of about 90 years. First, the promotion process used to be based only on the examination of a research thesis which included an oral examination and an assessment of the lecturing skills of the candidate. Then, the Higher Education Council, a national body, was established in 1981 and it was decided that scientometric data, such as the number of articles and citations, should be considered for evaluating research performance, in a manner similar to the American science system. At that time, there was still an oral examination. More recently, in 2016, the oral assessment was dropped from the promotion process and only scientometric data has been assessed since then.

In the below quote, a research manager from another institution explains that faculty members have teaching targets, but at the time of the interview, most of them were not evaluated on their research activities. He explains that he imported a research evaluation framework used in the United Kingdom (where he studied) and discussed it with several faculty members who agreed that such research targets were realistic goals to achieve. This framework, based on the number of publications, was first implemented in his department as a pilot project but ultimately it would be adopted across the different schools of the university:

> Before there were no [research] targets. There are teaching targets. The workload is defined in terms of teaching, but in terms of research, there was no system in place. So, I came up with this idea, inspired from the United Kingdom framework, like research targets and then I developed the new policy in which we have divided our faculty members into four different categories based on the number of courses and publications: *high teaching*, *balanced teaching*, *balanced research* and *high research* [...] Then I was designing this policy and I discussed it with different faculty members individually. I received their feedback and the majority of them agreed that this is a realistic target. As academicians, we have to develop our profiles. This is a pilot project. So, in the long run, this is beneficial for us and ultimately the university is going to implement it across the different schools in the next 2-3 years. They will have very strict research-related targets. (Respondent #11, Pakistan)

The four different categories are defined based on the teaching workload and the number of papers published in journals with impact factor (JIF) as follows in Table 1:

**Table 1. Profile Categories of faculty members based on their teaching and research targets per calendar year.**

| *Profile Category* | *Qualification* | *Teaching Target* | *Research Target* |
|---|---|---|---|
| *High teaching* | Non-PhD: Masters, Chartered Financial Analyst (CFA), and/or Association of Chartered Certified Accountants (ACCA) certificate | Six courses | One case study, a research grant, or a working paper. |
| *Balanced teaching* | PhD | Four courses | One paper (JIF) |
| *Balanced research* |  | Three courses | Two papers (JIF) |
| *High research* |  | Two courses | Three papers (JIF) |

From the previous quote, research activities seem to be normalized and rationalized. Academics are increasingly subject to quantitative and measurable outcomes that control requirements within new types of higher research systems (Burrows, 2012; Sauder & Espeland, 2009). Such transformations occur through the implementation of new incentives and policies as described in the quotes of this section.

In the next quote, the research manager explains that the quartiles of journals serve as indicators used for promotion purposes. They are used as global standards by the institution to set an evaluation framework. In this case, the research manager relies on the information extracted from the Journal Citations Reports to make a promotion decision. More explicitly, publishing 8 articles in Q1 journals allows the researcher to be promoted to the rank of Research Professor without being reviewed by peer-examiners and much faster than the usual process:

> If any researcher submits 8 articles for publication in Q1 journals and they are accepted, he/she will be promoted on a fast track. Fast track means it won't take 3 or 4 months to be promoted, he/she will be promoted in just 1 month. And the researcher's portfolio of publications will not be reviewed by examiners of the promotion committee. So, I recommend the researchers publish their work in Q1 journals. From where do we know this information? The only way is through Journal Citations Reports (JCR). I don't recommend any other website. I recommended that all researchers should know how to use JCR, how to find the quartile, and what are the differences between journals in Pharmacy and Chemistry. (Respondent #4, Egypt)

These different quotes show that research managers make practical decisions based on scientometric data. This section suggests that decision-making situations such as hiring, and promotion practices are closely related to the use of scientometric information. Hiring and promoting faculty members are done by using judgment devices, as research managers have to recommend someone or a group from a range of 'singularities' or entities with unique multidimensional qualities (Karpik, 2010). The authority of the research managers is exercised in various ways as previously mentioned.

*5.2 Funding allocation*

What I notice is that scientometric-based rules legitimate or inform decision-making. The research managers also draw on scientometrics to allocate funding. Such practice is described and discussed in this sub-section. In the quote that follows, the research manager explains that having access to a bibliometric database such as the Web of Science represents an advantage in providing some of the required information to allocate budgets to the researchers' groups. This budget allocation is done based on the number of publications indexed in the Web of Science over a specific period. The researcher manager does not need to rely anymore on the researchers to retrieve such information:

> There is an advantage if you have access to databases, like the Web of Science, you don't need to rely on researchers to check the production of their laboratory. Because we will allocate the budgets based on such information. We evaluate the scientific production over a period of 4 years of research structures from institutions based on certain indicators: published articles, books, patents, oral and written communications at conferences, national and international collaborations, organization of national and international conferences, etc.[2] (Respondent #8, Morocco)

Another example of a decision based on scientometric data is described in the below quote. The researcher manager participated in the writing of a scientometric report which aims at presenting an overview of the scientific publications of her institute. This report includes several elements such as the internal and external collaborations at the institute level. A decision was made to set a funding program focused on internal research projects to encourage the researchers to collaborate with internal colleagues on different topics:

> Recently, there was a need to have an "overview" of the scientific publications of the institute to see where we stand, where we publish the most, and which interactions we have internally/externally. We saw in the report that we wrote that there were not many internal collaborations although the research is very internationalized. So, the Management set up an internal research project funding program to encourage researchers to work internally and create bridges between different topics and prevent the teams to work in silos.[3] (Respondent #7, Tunisia)

The quotes presented in this section highlight the practical role of scientometric data in the decision-making processes of research managers. These quotes demonstrate the close relationship between funding allocation and the use of scientometrics. Research managers often rely on scientometrics as judgment devices to allocate funding, as they must assess a range of unique entities with multidimensional qualities (Karpik, 2010). As demonstrated by the quotes, scientometric indicators can provide information to support these judgments. Another example of such a decision-making context is illustrated in the latter quote where the way researchers collaborate is analyzed from a scientometric perspective and, as a result of this analysis, a specific funding project was set to encourage internal collaboration.

*5.3 University rankings*

Recently, there has also been an emphasis on quantitative indicators in science, where global rankings publish annual league tables for grading research and/or teaching. Knowledge of the bibliometrics industry by research managers includes the global university rankings, which have played a critical role in transforming higher education and science systems into a competition for students, reputation, and resources. Based on the methodologies of these rankings, some of their indicators become calculable. As a consequence, research managers analyze the research output of their institution from this ranking perspective. Then, as explained in the following quote, they may issue recommendations on several fronts such as setting new policies to award a Master's or a PhD degree based on the indexation of one's publication(s) in the Web of Science and Scopus:

> I submit a report to the management to highlight the strengths and weaknesses of our University once every three months and once every year using two main tools: SciVal and InCites. Because I know the US News and Shanghai rankings use Web of Science data and THE and QS use Scopus data. There is no big difference for us. But, for all Master' and PhD theses to be awarded, we recommend that the articles of the candidate should be published in Scopus and Web of Science, especially in the Faculty of Science, Medicine, Engineering … all science-related subjects. For Arts and Humanities subjects, papers are mainly published in Arabic. So, we just started to ask for the title/abstract/keywords to be available in English as well […] I never look at the rankings as a goal. I look at them as a tool to analyse and reach our goals. I need our university to be ranked in other rankings, for example, QS, US news, and Shanghai ranking in many different subject categories. I also want other subject categories to be cited

> like Humanities and Social Sciences. I like the THE and Shanghai rankings. But the rankers are very different, with different methodologies, and different key indicators … THE ranking is very different from Shanghai. Shanghai rankings are more research oriented than THE or QS. (Respondent #4, Egypt)

A requirement in terms of the language of publication is also mentioned for publications in Arts and Humanities, imposing the title, abstract and keywords of a manuscript to be written in English in addition to Arabic. This is a requirement for indexation in citation databases like the Web of Science. And such language requirement becomes an implicit global standard. However, the same manager is conscious of the differences between global universities rankings. They differ in terms of methodology but also in terms of research orientation. To a certain extent, the indexation of a journal and the publication in English form a set of different 'global standards' from the ranking perspective.

The global university rankings guide certain research policies in terms of the selection of publication venue and publication language but also in terms of financial incentives as explained in the quote that follows. This research manager clearly explains the rationale behind such a policy change. The change is influenced by the global university rankings, which use scientometric data to rank universities.

> We are responsible for understanding the different global rankings. And we try to work or assess our situation as a university based on the rankings. Then we introduce some recommendations to the Higher Management such as the change of internal policies. Egypt is now more interested in rankings. As an example of a change of the policy we did this year, we used to have a financial reward rule which was very weak and was not very attractive to researchers. We introduced citations into the award rule. We are referring to the 2 main databases: Web of Science and Scopus. […] Generally, our role is to set the right policy. And everyone tries to adapt to it and work with it […] It would be an overview guideline, not a detailed one. We would tell them, if you publish in these journals, for example, Q1 journals, you would get a 10% extra financial reward. If you publish in Nature or Science, you will get 20% extra. And these are the categories of the journals. If a journal is indexed and it has an Impact Factor, then you would get these points. (Respondent #5, Egypt)

Countries and individual institutions show an interest in such global university rankings and play along the set of rules used by ranking agencies to improve their ranks worldwide. Consequently, new financial incentives are introduced locally as the 'right' policies to direct the research output of researchers as per the various global rankings rules. These global rules are adopted and adapted locally by the research managers. The process of glocalization is manifested through the implementation of publication guidelines, which incentivize researchers financially based on the venue of their publications. The level of the financial reward is decided based on the name or the 'brand' of the journal, its indexation, and its Journal Impact Factor quartile.

The detailed use of scientometric data shown in this section points to the fact that such data is used to set new policies in various contexts. These contexts include hiring and promotion practices, setting financial incentives and research publications targets, funding allocation, and university rankings.

## 6. Discussion

In this study, I have argued that scientometric data and indicators direct the transformation process of science systems in MENA. The results provide a better understanding of how research institutions in this specific region adopt scientometrics as 'global norms' and adapt them to alter local research systems. This study contributes to recent debates which have focused on how research is funded, conducted and assessed. Over the past 30 years, the organizational capacity of academic institutions has grown in importance as a result of the rise of assessments in science systems. Institutional management has become more complex (Simon & Knie, 2013), and research evaluation now plays a significant role in this complexity (Whitley & Gläser, 2007).

The findings of this study show that scientometric indicators are adopted as 'global standards' (Paradeise, 2016; Paradeise & Thoenig, 2013) by research managers in MENA. These indicators include simple and more technical indicators, such as the citation counts, the journals' impact factors and the journals' quartiles. The recent developments in technology and scientometrics made these indicators widely available. This study also shows that scientometric indicators are adopted at various levels within research institutions. Scientometric indicators are communicated to the researchers through workshops in various contexts such as the selection of the publishing venue, or promotion. When sharing such information, research managers may face some challenges and resistance from researchers who have their own opinions. This negotiation follows a glocalization process (Robertson, 2012) in which research managers communicate the value of scientometrics in their own way. In that sense, scientometric indicators are adopted as standards or norms but also adapted locally by the different research stakeholders. This glocalization process occurs partly through the development of new decision-making processes by research managers who draw on scientometric data as judgment devices for decision-making purposes. Hiring, promoting and allocating budget are done by using judgment devices, as research managers have to recommend someone or a group from a range of entities with unique multidimensional qualities or 'singularities' (Karpik, 2010).

The use of scientometric data and indicators by research managers to set new policies contributes also to the glocalization process. This process involves adapting global standards to local contexts and creating glocal standards that reflect local needs and priorities. This implies the setting of new scientometric-based rules and organization goals which constitute a very direct form of implementation of scientometrics in science systems. As a result, the use of scientometric data and indicators in decision-making processes, such as setting new policies, becomes an essential aspect of the glocalization process. The integration of scientometrics into local research management practices aims at facilitating the adoption of new policies that reflect both global and local perspectives. The use of scientometric data and indicators by research managers leads to the creation of new rules and policies based on scientometrics. For instance, research managers may establish scientometric-based guidelines for publishing in high-impact journals and set financial incentives for researchers who meet these targets. Similarly, research managers may use them to determine promotion and tenure decisions for faculty. These scientometric rules create a direct and tangible influence of scientometrics on science systems. The use of scientometric data and indicators is embedded in organizational processes which tend to copy the systems used in countries such as the United States and the United Kingdom. Research managers may set research publication targets, new promotion processes, as well as policies to award a Master or a PhD degree. Consequently, researchers adapt themselves to these new scientometric rules which create new science systems.

The results of this study highlight that several MENA research institutions are relying more and more on a set of standards that are established externally and adapted internally to define and assess academic quality (Paradeise, 2016; Paradeise & Thoenig, 2013). Such a situation occurs in the context of internationalization, which is related to the university's rank on global ranking systems as shown by Hazelkorn (2015, 2018). In such rankings, the research-related metrics influence a university's position and impact national science systems on various fronts. Research governance is increasing and research evaluation takes a prominent role in such change (Whitley & Gläser, 2007). Indeed, the findings of this study show that research evaluation manages various sources of influence, control, and governance at different levels: faculty hiring, faculty promotion, research funding, publishing, collaboration, decision-making and policy development.

In that context, the *More Than Our Rank* initiative has been developed in response to some of the problematic effects global university rankings have. This initiative aims also at highlighting the various ways universities serve the world that are not reflected in rankings. Several initiatives, such as *the San Francisco Declaration on Research Assessment (DORA)*, *The Leiden Manifesto for research metrics,* and the *Coalition for Advancing Research Assessment (COARA)* have all reflected on the role of metrics in evaluation frameworks. These initiatives have the potential to change the way scientometrics is used in specific countries of the world. Many government entities and research institutions have already designed and implemented richer frameworks to assess research. This suggests that 'global standards' are also evolving because of these initiatives. Therefore, MENA countries may be adopting scientometrics as 'global standards' from the past rather than the new 'global standards' that may emerge from these recent initiatives. It is critical to understand that 'global standards' are indeed dynamic. Future research might seek to study the dynamic nature of 'global standards' in research assessment.

## 7. Acknowledgments
I would like to thank Thomas Franssen for his guidance on this project and I am grateful to Ludo Waltman for his valuable comments and suggestions. Lastly, I appreciate the engagement of the two anonymous reviewers for dedicating their time and expertise to the review process.

## 8. Funding and competing interests
No funding was received for this study. The author is an employee of Clarivate Analytics, the provider of Web of Science, Journal Citations Reports, Essential Science Indicators and InCites Benchmarking & Analytics mentioned in this study.

## References
Agyemang, G., & Broadbent, J. (2015). Management control systems and research management in universities. *Accounting, Auditing & Accountability Journal*, *28*(7), 1018-1046. https://doi.org/10.1108/aaaj-11-2013-1531

Anninos, L. N. (2014). Research performance evaluation: some critical thoughts on standard bibliometric indicators. *Studies in Higher Education*, *39*(9), 1542-1561. https://doi.org/10.1080/03075079.2013.801429

Basu, A. (2006). Using ISI's' Highly Cited Researchers' to obtain a country level indicator of citation excellence. *Scientometrics*, *68*(3), 361-375. https://doi.org/10.1007/s11192-006-0117-x

Berman, E. P. (2011). *Creating the market university: How academic science became an economic engine*. Princeton University Press.

Brisset-Sillon, C. (1997). Universités publiques aux États-Unis: Une autonomie sous tutelle. *Universités publiques aux États-Unis*, 1-300.

Brunsson, N., & Jacobsson, B. (2002). *A world of standards*. Oxford University Press. https://doi.org/10.1093/acprof:oso/9780199256952.001.0001

Brunsson, N., Rasche, A., & Seidl, D. (2012). The dynamics of standardization: Three perspectives on standards in organization studies. *Organization Studies*, *33*(5-6), 613-632. https://doi.org/10.1177/0170840612450120

Burrows, R. (2012). Living with the h-index? Metric assemblages in the contemporary academy. *The sociological review*, *60*(2), 355-372. https://doi.org/10.1111/j.1467-954X.2012.02077.x

de Rijcke, S., Wouters, P. F., Rushforth, A. D., Franssen, T. P., & Hammarfelt, B. (2016). Evaluation practices and effects of indicator use—a literature review. *Research Evaluation*, *25*(2), 161-169. https://doi.org/10.1093/reseval/rvv038

Dearlove, J. (1997). The academic labour process: from collegiality and professionalism to managerialism and proletarianisation? *Higher Education Review*, *30*(1), 56.

Durand, T., & Dameron, S. (2011). Where have all the business schools gone? *British Journal of Management*, *22*(3), 559-563. https://doi.org/10.1111/j.1467-8551.2011.00775.x

Ellegaard, O., & Wallin, J. A. (2015). The bibliometric analysis of scholarly production: How great is the impact? *Scientometrics*, *105*(3), 1809-1831.

Elo, S., & Kyngäs, H. (2008). The qualitative content analysis process. *Journal of Advanced Nursing*, *62*(1), 107-115. https://doi.org/10.1111/j.1365-2648.2007.04569.x

Franssen, T., & Wouters, P. (2019). Science and its significant other: Representing the humanities in bibliometric scholarship. *Journal of the Association for Information Science and Technology*, *70*(10), 1124-1137. https://doi.org/10.1002/asi.24206

Garfield, E. (2009). From the science of science to Scientometrics visualizing the history of science with HistCite software. *Journal of Informetrics*, *3*(3), 173-179. https://doi.org/10.1016/j.joi.2009.03.009

Garfield, E. (2012). A Century of Citation Indexing. *Collnet Journal of Scientometrics and Information Management*, *6*(1), 1-6. https://doi.org/10.1080/09737766.2012.10700919

Geuna, A., & Martin, B. R. (2003). University Research Evaluation and Funding: An International Comparison. *Minerva*, *41*(4), 277-304. https://doi.org/10.1023/b:mine.0000005155.70870.bd

Hazelkorn, E. (2015). *Rankings and the reshaping of higher education: The battle for world-class excellence*. Springer. https://doi.org/10.1057/9781137446671

Hazelkorn, E. (2018). Reshaping the world order of higher education: the role and impact of rankings on national and global systems. *Policy Reviews in Higher Education*, *2*(1), 4-31. https://doi.org/10.1080/23322969.2018.1424562

Hicks, D. (2012). Performance-based university research funding systems. *Research Policy*, *41*(2), 251-261. https://doi.org/j.respol.2011.09.007

Hirsch, J. E. (2005). An index to quantify an individual's scientific research output. *Proceedings of the National academy of Sciences*, *102*(46), 16569-16572. https://www.ncbi.nlm.nih.gov/pmc/articles/PMC1283832/pdf/pnas-0507655102.pdf

Jiménez-Contreras, E., de Moya Anegón, F., & López-Cózar, E. D. (2003). The evolution of research activity in Spain: The impact of the National Commission for the Evaluation of Research Activity (CNEAI). *Research Policy*, *32*(1), 123-142. https://doi.org/10.1016/S0048-7333(02)00008-2

Karpik, L. (2010). *The economics of singularities*. Princeton University Press, Princeton. https://doi.org/10.2307/j.ctv1zm2v2n

Krücken, G., & Meier, F. (2006). Turning the university into an organizational actor. *Globalization and organization: World society and organizational change*, 241-257.

Lahtinen, E., Koskinen-Ollonqvist, P., Rouvinen-Wilenius, P., Tuominen, P., & Mittelmark, M. B. (2005). The development of quality criteria for research: a Finnish approach. *Health Promotion International*, *20*(3), 306-315. https://doi.org/10.1093/heapro/dai008

Langfeldt, L., Nedeva, M., Sörlin, S., & Thomas, D. A. (2020). Co-existing notions of research quality: A framework to study context-specific understandings of good research. *Minerva*, *58*(1), 115-137. https://doi.org/10.1007/s11024-019-09385-2

Margolis, J. (1967). Citation Indexing and Evaluation of Scientific Papers: The spread of influence in populations of scientific papers may become a subject for quantitative analysis. *Science*, *155*(3767), 1213-1219. https://doi.org/10.1126/science.155.3767.1213

Moed, H., De Bruin, R., Nederhof, A., Van Raan, A., Tijssen, R., Schermer, L., & Removille, J. (1992). State of the art bibliometric macro-indicators(an overview of demand and supply). *EUR(Luxembourg)*.

Moed, H. F., Burger, W. J. M., Frankfort, J. G., & Van Raan, A. F. J. (1985). The use of bibliometric data for the measurement of university research performance. *Research Policy*, *14*(3), 131-149. https://doi.org/10.1016/0048-7333(85)90012-5

Morris, H., Harvey, C., Kelly, A., & Rowlinson, M. (2011). Food for Thought? A Rejoinder on Peer-review and RAE2008 Evidence. *Accounting Education*, *20*(6), 561-573. https://doi.org/10.1080/09639284.2011.634215

Musselin, C. (1996). Les marchés du travail universitaires, comme économie de la qualité. *Revue française de sociologie*, 189-207.

Musselin, C. (2005). *Le marché des universitaires: France, Allemagne, États-Unis*. Presses de Sciences Po.

Musselin, C. (2013). *The long march of French universities*. Routledge.

Narin, F. (1976). *Evaluative bibliometrics: The use of publication and citation analysis in the evaluation of scientific activity*. Computer Horizons Cherry Hill, NJ.

Narin, F., & Hamilton, K. S. (1996). Bibliometric performance measures. *Scientometrics*, *36*(3), 293-310. https://doi.org/10.1007/bf02129596

Osterloh, M. (2010). Governance by numbers. Does it really work in research? *Analyse & Kritik*, *32*(2), 267-283.

Paradeise, C. (2016). *In search of academic quality*. Springer. https://doi.org/10.1057/9781137298294

Paradeise, C., & Thoenig, J.-C. (2013). Academic Institutions in Search of Quality: Local Orders and Global Standards. *Organization Studies*, *34*(2), 189-218. https://doi.org/10.1177/0170840612473550

Peterson, M. W. (2007). The study of colleges and universities as organizations. *Sociology of higher education: Contributions and their contexts*, 147-186.

Porter, T. M. (1996). Trust in numbers. In *Trust in Numbers*. Princeton University Press. https://doi.org/10.1515/9781400821617

Renaut, A. (1995). *Les révolutions de l'université: essai sur la modernisation de la culture*. FeniXX.

Robertson, R. (2012). Globalisation or glocalisation? *The Journal of International Communication*, *18*(2), 191-208. https://doi.org/10.1080/13216597.2012.709925

Sauder, M., & Espeland, W. N. (2009). The discipline of rankings: Tight coupling and organizational change. *American sociological review*, *74*(1), 63-82. https://doi.org/10.1177/000312240907400104

Simon, D., & Knie, A. (2013). Can evaluation contribute to the organizational development of academic institutions? An international comparison. *Evaluation*, *19*(4), 402-418. https://doi.org/10.1177/1356389013505806

Siow, A. (1995). *The organization of the market for professors*.

Sivertsen, G. (2018). The Norwegian Model in Norway. *Journal of Data and Information Science*, *3*(4), 3-19. https://doi.org/10.2478/jdis-2018-0017

Thelwall, M., Kousha, K., Wouters, P., Waltman, L., de Rijcke, S., Rushforth, A., & Franssen, T. (2015). *The metric tide: Literature review*. https://doi.org/10.13140/RG.2.1.5066.3520

Thomas, D. A., Nedeva, M., Tirado, M. M., & Jacob, M. (2020). Changing research on research evaluation: A critical literature review to revisit the agenda. *Research Evaluation*, *29*(3), 275-288. https://doi.org/10.1093/reseval/rvaa008

Van Vught, F. A. (1995). Policy models and policy instruments in higher education: The effects of governmental policy-making on the innovative behaviour of higher education institutions.

Weingart, P. (2005). Impact of bibliometrics upon the science system: Inadvertent consequences? *Scientometrics*, *62*(1), 117-131. https://doi.org/10.1007/s11192-005-0007-7

Werner, R. (2015). The focus on bibliometrics makes papers less useful. *Nature News*, *517*(7534), 245. https://doi.org/10.1038/517245a

Whitley, R., & Gläser, J. (2007). The changing governance of the sciences. *Sociology of the sciences yearbook*, *26*. https://doi.org/10.1007/978-1-4020-6746-4_1

**Notes**

[1] Original (French) : Un autre challenge pour nous est de savoir comment convaincre les chercheurs que les indicateurs qui sont déterminés par les grosses bases de données sont les bons. J'ai la moitié de mes chercheurs qui ne croient pas au facteur d'impact. Ils me disent que 'ça ce n'est pas important. Je ne vais pas travailler sous la pression'. Ça c'est un challenge, je dis à chaque fois que c'est dur de leur prouver que c'est la meilleure solution etc… bien que nous sommes classés dans le classement national ou dans les classements mondiaux, pour moi c'est dur de faire comprendre que c'est important. Pour eux une publication c'est bon, elle n'est pas bonne ou mauvaise car le facteur d'impact est de tant ou de tant.

[2] Original (French) : Il y a un avantage si on a accès aux bases de données, comme le Web of Science, on n'a pas besoin de faire recours aux déclarations des professeurs-chercheurs pour vérifier la production d'un laboratoire etc… Car c'est à partir de là qu'on va repartir les budgets. Donc on évalue la production scientifique (sur une période de 4 ans) des structures de recherche issues des établissements en se basant sur certains indicateurs à savoir : articles publiés, livres, brevets, communications orales et écrites dans des congrès, collaborations nationales et internationales, organisation des colloques nationales et internationales, etc…

[3] Original (French): Dernièrement, il n'y a pas très longtemps, il y avait un besoin d'avoir un 'overview', un suivi des publications scientifiques de l'institut, pour voir ou ce qu'on est, là où on publie le plus, quelles sont les interactions qu'on a en interne/externe. On a vu dans ce rapport qu'on a rédigé qu'il n'y avait pas beaucoup de collaborations en interne. Bien que la recherche soit très internationalisée. Donc ils ont mis en place un programme de financement de projets de recherche interne pour inciter les chercheurs à travailler en interne et créer des passerelles entre différentes thématiques pour éviter le cloisonnement des équipes entre elles.